\documentclass[aps,prc,letterpaper,11pt,twoside,tightenlines,nofootinbib,showpacs,preprint]{revtex4}
\usepackage{graphicx}
\usepackage[sort&compress]{natbib}
\begin{document}

\newcommand{\be}{\begin{equation}}
\newcommand{\ee}{\end{equation}}
\newcommand{\bea}{\begin{eqnarray}}
\newcommand{\eea}{\end{eqnarray}}
\newcommand{\bef}{\begin{figure}}
\newcommand{\eef}{\end{figure}}
\newcommand{\bce}{\begin{center}}
\newcommand{\ece}{\end{center}}

\title{The $\pi\rho$ Cloud Contribution to the $\omega$ Width in Nuclear Matter}
\author{D.~Cabrera}
\affiliation{Institute for Theoretical Physics, Frankfurt University, 
60438 Frankfurt am Main, Germany}
\affiliation{Frankfurt Institute for Advanced Studies, Frankfurt University, 60438 Frankfurt am Main, Germany}
\author{R.~Rapp}
\affiliation{Cyclotron Institute and Department of Physics~\&~Astronomy, Texas A\&M
University, College Station, Texas 77843-3366, U.S.A.}

\date{\today}

\begin{abstract}
The width of the $\omega$ meson in cold nuclear matter is computed in a hadronic 
many-body approach, focusing on a detailed treatment of the medium modifications of 
intermediate $\pi\rho$ states. The $\pi$ and $\rho$ propagators are dressed by their 
selfenergies in nuclear matter taken from previously constrained many-body 
calculations. The pion selfenergy includes $Nh$ and 
$\Delta h$ excitations with short-range correlations, while the $\rho$ selfenergy 
incorporates the same dressing of its $2\pi$ cloud with a full 3-momentum dependence and 
vertex corrections, as well as direct resonance-hole excitations; both contributions 
were quantitatively fit to total photo-absorption spectra and 
$\pi N\to\rho N$ scattering. Our calculations account for in-medium decays of 
type $\omega N\to \pi N^{(*)},~\pi\pi N(\Delta)$, and 2-body absorptions 
$\omega NN \to NN^{(*)},~\pi NN $. This causes deviations of the in-medium 
$\omega$ width from a linear behavior in density, with important contributions
from spacelike $\rho$ propagators. The $\omega$ width from the $\rho\pi$ cloud may reach
up to 200~MeV at normal nuclear matter density, with a moderate 3-momentum dependence.
This largely resolves the discrepancy of linear $T$-$\varrho$ approximations with the 
values deduced from nuclear photoproduction measurements.
\end{abstract}

\maketitle

\section{Introduction}
\label{sec:intro}

The low-mass vector mesons $\rho$, $\omega$ and $\phi$ play a special role in the 
study of hot and dense nuclear matter, as their dilepton decay channel ($l^+l^-$) 
provides a pristine window on their in-medium properties. This feature has been 
extensively and 
successfully exploited in the measurement of dilepton spectra in heavy-ion 
collisions~\cite{Tserruya:2009zt,Specht:2010xu,Geurts:2012rv}. In these reactions, 
the thermal emission of low-mass dileptons is dominated by the 
$\rho$ meson, due to its much larger dilepton width compared to the $\omega$, 
$\Gamma_{\rho\to ll}\simeq 10~\Gamma_{\omega\to ll}$. Dilepton data from the SPS and 
RHIC can now be consistently understood by a strong broadening (``melting") 
of the $\rho$ meson, as computed from hadronic many-body theory in the hot and 
dense system~\cite{Rapp:2009yu,Rapp:2013ema}. This approach also yields a good 
description~\cite{Leupold:2009kz,Riek:2010gz} of the $\rho$ broadening observed in 
nuclear photoproduction, if the data are corrected with absolute background 
determination~\cite{Huber:2003pu,Wood:2008ee}. As a further test of the validity and 
generality of the hadronic in-medium approach, the $\omega$ meson, as the isospin zero 
pendant of the $\rho$, is a natural candidate. 

The small dilepton decay width of the $\omega$ led the CB-TAPS collaboration to
pursue the $\pi^0\gamma$ decay channel in photon-induced production off nuclei.  
Early results for invariant-mass spectra reported significant downward mass 
shifts~\cite{Trnka:2005ey}, seemingly in line with proton-induced 
dilepton production off nuclei~\cite{Naruki:2005kd}. 
However, with improved background determination these results were not 
confirmed~\cite{Kaskulov:2006zc,Nanova:2010sy}, leaving no evidence for a mass drop.  
As an alternative method, absorption measurements have been performed for $\phi$
and $\omega$ mesons in $e^+e^-$~\cite{Ishikawa:2004id,Wood:2010ei} and 
$\pi^0\gamma$~\cite{Kotulla:2008aa} channels. These data are not directly
sensitive to possible mass shifts, but they can be used to assess the in-medium 
(absorptive) widths. For both $\phi$ and $\omega$, large in-medium widths have been 
deduced, e.g., $\Gamma_\omega^{\rm med}\simeq$~130-150\,MeV~\cite{Kotulla:2008aa}, or even
above 200\,MeV~\cite{Wood:2010ei}, for the $\omega$ at normal 
nuclear matter density. These values exceed the free $\omega$ width 
by a factor of $\sim$20, posing a challenge for theoretical 
models~\cite{Klingl:1997kf,Post:2000rf,Lutz:2001mi,Zschocke:2002mp,Martell:2004gt,Muehlich:2006nn,Eichstaedt:2007zp,Rodrigues:2011zzb,Ghosh:2012sa}.


Most of the calculations thus far are based on the so-called $T$-$\varrho$ approximation, 
where the in-medium $\omega$ selfenergy is computed from the vacuum scattering amplitude
and therefore depends linearly on nuclear density, $\varrho_N$ (see, however, 
Refs.~\cite{Wachs:2000,Riek:2004kx}).
In the present work we go beyond this approximation 
by simultaneously dressing the $\pi$ and $\rho$ propagators in the $\pi\rho$ loop of 
the $\omega$ selfenergy.
In the vacuum, the $\omega$ decay into $\pi\rho$ has a nominal threshold of 
$m_\pi+m_\rho\simeq910$\,MeV and only proceeds through the low-mass tail of the 
$\rho$ resonance, which is suppressed and possibly responsible for the small 
width of $\Gamma_{\omega\to 3\pi}\simeq$~7.5\,MeV. 
A broadening of the $\rho$ in the medium enhances this decay channel,
further augmented if the pion is dressed as well.
This is a key point we aim to convey and elaborate quantitatively in this paper 
by utilizing realistic in-medium $\pi$ and $\rho$ propagators.

Our paper is organized as follows. In Sec.~\ref{sec:self} we 
set up the $\omega\to\pi\rho$ selfenergy in vacuum (Sec.~\ref{ssec:vac}) and 
discuss the implementation of the $\pi$ and 
$\rho$ propagators in nuclear matter (Sec.~\ref{ssec:med}). In Sec.~\ref{sec:results} we 
quantitatively evaluate the consequences of the in-medium propagators on the density 
and 3-momentum dependence of the $\omega$ width. We summarize and give an
outlook in Sec.~\ref{sec:sum}.

\section{$\omega$ Selfenergy}
\label{sec:self}

\subsection{$\omega$ Width in Vacuum}
\label{ssec:vac}
In vacuum we describe the coupling of the $\omega$ to a pion and a $\rho$ meson 
with the chiral anomalous interaction Lagrangian 
introduced, e.g., in the work by Schechter {\it et al.}~\cite{Jain:1987sz},  
\begin{equation}
{\cal L}_{\omega\rho\pi}^{\rm int} = g_{\omega\rho\pi} \epsilon_{\mu\nu\sigma\tau}
        \partial^\mu \omega^\nu \partial^\sigma \vec{\rho}^{\,\tau} \cdot \vec\pi \ .
\label{L}
\end{equation}
The value of the coupling constant, $g_{\omega\rho\pi}$, determines the partial decay
width $\Gamma_{\omega\to\rho\pi}$ and will be discussed below. 
A straightforward application of Feynman rules for the $\pi\rho$ loop yields 
the polarization-averaged selfenergy of an $\omega$ of 4-momentum $P=(P^0,\vec P)$ as
\bea
-i \, \Pi_{\omega}(P) &=& IF \,
 \frac{1}{3} \, \sum_{\lambda,\,\delta}
\epsilon_{\lambda}^{\nu}(P) 
\epsilon_{\delta}^{\nu'}(P) \
\ i g_{\omega\rho\pi}  \ i g_{\omega\rho\pi} \ 
\varepsilon_{\mu\nu\alpha\beta} \  \varepsilon_{\mu'\nu'\alpha'\beta'}
\nonumber \\
&&\times
\int \frac{d^4q}{(2\pi)^4} \  P^{\mu} q^{\alpha} \ P^{\mu'} q^{\alpha'} \
i D_{\rho}^{\beta \beta'}(q) \ i D_{\pi}(P-q) \ ,  
\label{Pi_omg-pol}
\eea
where the isospin factor $IF$=3 accounts for the different $\pi\rho$ charge states.
Using standard representations of the polarization sum and of the spin-1 $\rho$ propagator, 
$D_{\rho}^{\beta \beta'}$, which we decompose in transverse (T) and longitudinal (L) 
modes~\cite{Urban:1998eg}, one finds
\begin{equation}
-i \, \Pi_{\omega}(P) = -\frac{4}{3} IF g_{\omega\rho\pi}^2 \int \frac{d^4q}{(2\pi)^4} \ 
D_{\pi}(P-q) \  \lbrace 
v_1(q,P) \,D^T_{\rho}(q) 
+
v_2(q,P) \, \lbrack D^T_{\rho}(q) - D^L_{\rho}(q) \rbrack
\rbrace
 \ 
\label{Pi-omg}
\end{equation}
where $D_{\pi}(P-q)=1/[(P-q)^2-m_\pi^2-\Pi_{\pi}]$ and 
$D^{T,L}_{\rho}(q) =1/[q^2-M_\rho^2-\Pi^{T,L}_{\rho}]$ are the scalar parts of the meson
propagators with complex selfenergies. The two vertex functions arise from the Lorentz 
contractions with the T and L projectors of the $\rho$ propagator, 
$v_1(q,P)=P^2 q^2 - (P q)^2$ and 
$v_2(q,P)=q^2\left(\vec{P}\,^2 - \vec{P}\cdot\vec{q}/\vec{q}\,^2\right)/2$. 
The above expression is valid both in vacuum and in medium and incorporates the 
$\omega$ 3-momentum dependence.  Using the Lehmamn representation for the propagators 
one finds
\begin{eqnarray}
\label{Pi-omg-spec}
\Pi_{\omega}(P) &=& - 2\ \frac{4}{3} IF g_{\omega\rho\pi}^2 
\int_0^{\infty} d\omega \int_0^{\infty} d\omega' \frac{\omega+\omega'}{(P^0)^2-(\omega+\omega')^2+i\eta} \nonumber \\
&&\times \int \frac{d^3q}{(2\pi)^3} S_{\pi}(\omega',\vec{P}-\vec{q}\,)
\lbrace 
v_1(q,P) S^T_{\rho}(q) 
+ v_2(q,P) \lbrack S^T_{\rho}(q) - S^L_{\rho}(q) \rbrack
\rbrace_{q^0=\omega}
\
\end{eqnarray}
with $S^{T,L}_{\rho}=-\frac{1}{\pi}\textrm{Im}D^{T,L}_{\rho}$, $S_{\pi}=-\frac{1}{\pi}\textrm{Im}D_{\pi}$ denoting the $\rho$ and $\pi$ spectral 
functions, respectively.
%
The $\omega$ width follows from the imaginary part of the selfenergy as
$\Gamma_{\omega\to\rho\pi}(P) = - \textrm{Im}\, \Pi_{\omega}(P)/P^0$.
In vacuum, free spectral functions for the pion and the $\rho$ meson
are utilized, 
\bea
\label{S-vac}
S_{\pi}^{\rm vac}(\omega', \vec{q}\,)=\delta(\omega'^2-\vec{q}\,^2-m_{\pi}^2) \ ,  \quad
S_{\rho}^{\rm vac}(\omega,\vec{q}) = -\frac{1}{\pi} 
\frac{\textrm{Im}\,\Pi^{\rm vac}_{\rho\pi\pi}(q^2)}
{|\omega^2-\vec{q}\,^2-M_{\rho}^2-\Pi^{\rm vac}_{\rho\pi\pi}(q^2)|^2} \ .
\eea
The $\rho\to\pi\pi$ selfenergy is often approximated by reabsorbing the real part 
into the physical 
$\rho$ mass, $m_\rho^2\equiv M_\rho^2-\textrm{Re}\,\Pi^{\rm vac}_{\rho\pi\pi}$, 
and an imaginary part
\bea
\label{Grho-vac}
\textrm{Im}\, \Pi^{\rm vac}_{\rho\pi\pi}(q^2) = -\frac{g_{\rho\pi\pi}^2}{48\pi \sqrt{q^2}} 
(q^2-4 m_{\pi}^2)^{\frac{3}{2}} \ \Theta(q^2-4 m_{\pi}^2)  \ 
\eea
with $g_{\rho\pi\pi}$$\simeq$6 to obtain $\Gamma_{\rho\to\pi\pi} = 
-\textrm{Im}\, \Pi^{\rm vac}_{\rho\pi\pi}(q^2=m_{\rho}^2) / m_{\rho}\simeq150$\,MeV.
Here, we use the microscopic vacuum spectral function underlying our in-medium 
model~\cite{Urban:1998eg}, which describes the low-mass tail of the $\rho$ 
resonance more accurately, incorporating an energy dependence of 
$\textrm{Re}\,\Pi^{\rm vac}_{\rho\pi\pi}$.
With $g_{\omega\rho\pi}=1.9/f_{\pi}$ 
($f_{\pi}$=92\,MeV)~\cite{Jain:1987sz,Klingl:1996by}, one obtains
$\Gamma_{\omega\to\rho\pi}=3.6$\,MeV, 
i.e., about 1/2 of the total 3$\pi$ width (2/3 when including interference 
effects~\cite{Gudino:2011ri}). Using a schematic Breit-Wigner $\rho$ spectral function, 
$\Gamma_{\omega\to\rho\pi}(m_\omega)$ is reduced by approximately 30\%. In 
Ref.~\cite{Gudino:2011ri} the partial $\pi\rho$ width was found to be 2.8\,MeV. Rescaling
our $g_{\omega\rho\pi}$ to obtain that value would entail an according 22\% reduction
of our in-medium widths reported below. Some of this would be recovered by medium 
effects of the accompanying increase in the direct 3$\pi$ channel.   


\subsection{$\rho$ and $\pi$ Propagators in Nuclear Matter}
\label{ssec:med}
Before proceeding to calculate the $\omega$ meson width in nuclear matter 
caused by the dressing of the propagators in the $\pi\rho$ loop, 
$\Gamma_{\omega\to\pi\rho}^{\rm med}$, two comments are in order. 


We first note that the unnatural-parity coupling in the $\omega\rho\pi$ Lagrangian 
(\ref{L}) implies transversality of any contribution to the $\omega$ selfenergy with 
at least one $\omega\rho\pi$ vertex with an external $\omega$~\cite{Wachs:2000}. 
Thus, in-medium vertex corrections, as required to ensure transversality for the
pion cloud of the $\rho$ meson~\cite{Chanfray:1993ue,Herrmann:1993za,Urban:1998eg}
(or chiral symmetry in the $\sigma$ channel~\cite{Chiang:1997di}), are not dictated
here, but correspond to contributions to
$\omega N\to \pi N,\,\pi\pi N$ scattering unrelated to the anomalous decay process.
We will not include these in the present work.

Second, at finite 3-momentum relative to the nuclear medium, the $\rho$ propagator 
splits into transverse and longitudinal modes. At $\vec P=\vec{0}$, the $\omega$ selfenergy 
only depends on the transverse modes of the $\rho$, since the vertex
function $v_2$ in Eq.~(\ref{Pi-omg}) vanishes. However, for $\vec P\ne \vec{0}$, $v_2$
becomes finite and proportional to $S^T_{\rho}-S^L_{\rho}$. This contribution turns
out to be appreciable due to the splitting of the in-medium T and L modes of the 
$\rho$~\cite{Urban:1998eg} within the kinematics of the $\omega\to\rho\pi$ decay.  

Let us turn to briefly reviewing the main ingredients to the evaluation
of $\Gamma_{\omega\to\pi\rho}^{\rm med}$ from Eq.~(\ref{Pi-omg-spec}), which 
are the microscopic calculations of the in-medium pion and $\rho$ propagators.


The pion spectral function is evaluated with standard $P$-wave nucleon-hole ($NN^{-1}$) 
and Delta-hole ($\Delta N^{-1}$) excitations~\cite{Oset:1981ih,Migdal:1990vm}. 
The corresponding irreducible $P$-wave pion self-energy,
\be
\label{piself}
\Pi_{\pi} (q^0,\vec{q};\varrho) =
\frac{\left( \frac{f_N}{m_{\pi}} \right) ^2
F_{\pi}(\vec{q}\,^2) \, \vec{q}\,^2 \,
\left[ U_{NN}  + U_{\Delta N}
-(g'_{11}-2g'_{12}+g'_{22})U_{NN} U_{\Delta N}
\right]}
{1 - \left( \frac{f_N}{m_{\pi}} \right) ^2 \, 
\left[ g'_{11}U_{NN} 
+ g'_{22}U_{\Delta N} 
-(g'_{11}g'_{22}-g_{12}'^2) U_{NN} 
U_{\Delta N}
 \right]}
\,\,\, ,
\ee
is given by the Lindhard functions $U_\alpha$ for the loop diagrams~\cite{Oset:1989ey}; 
they include transitions between the two channels through short-range correlations 
represented by Migdal parameters $g'$. The $\pi NN$ and $\pi N \Delta$ coupling 
constants, $f_N \simeq 1$ and  $f_{\Delta}/f_N \simeq 2.13$ (absorbed in the definition 
of $U_{\Delta N}$), are determined from pion-nucleon and pion-nucleus reactions. 
Finite-size effects on the $\pi NN$ and $\pi N \Delta$ vertices are simulated via 
hadronic monopole form factors, 
\be
F_{\pi}(\vec{q}\,^2) = \Lambda_{\pi}^2 /
(\Lambda_{\pi}^2 + \vec{q}\,^2) \ .
\label{FpiNN}
\ee
Consistency with our model for the in-medium $\rho$ discussed below dictates 
a soft cutoff, $\Lambda_{\pi}$=0.3\,GeV, following from constraints of 
$\pi N\to\rho N$ scattering data and the non-resonant continuum in nuclear 
photo-absorption~\cite{Rapp:1999ej} (e.g., with $\Lambda_{\pi}$=0.5\,GeV one 
overestimates the measured $\pi N\to\rho N$ cross section by a factor of $\sim$2). 
Especially the former probe similar kinematics 
of the virtual $\pi NN$ vertex as figuring into  
$\omega N\to \rho N$ processes. The Migdal parameters are
$g'_{11}=0.6$ and $g'_{12}=g'_{22}=0.2$.

The in-medium $\rho$ spectral function is taken from Refs.~\cite{Rapp:1997ei,Urban:1998eg},
which start from a realistic description of the $\rho$ in free space (reproducing $P$-wave 
$\pi\pi$ scattering and the pion electromagnetic form factor). The selfenergy in nuclear 
matter contains two components: pisobars ($NN^{-1}$, $\Delta N^{-1}$) in 
the two-pion cloud, $\Pi_{\rho\pi\pi}$, and direct baryon resonance excitations 
in $\rho N$ scattering, $\Pi_{\rho BN^{-1}}$ (``$\rho$-sobars").
The latter have been evaluated using effective Lagrangians in hadronic many-body theory 
(in analogy to the pion)~\cite{Urban:1998eg,Urban:1999im,Rapp:1999us}, including ca.~10 
baryonic resonances. In $\Pi_{\rho\pi\pi}$, the in-medium pion propagator described 
above is supplemented with vertex corrections to preserve the Ward-Takahashi identities 
of the $\rho$ propagator; it extends to finite 3-momentum of the $\rho$ which is essential 
for the $\pi\rho$ loop in $\Pi_\omega$. The total $\rho$ selfenergy is quantitatively 
constrained by nuclear photo-absorption and $\pi N\to\rho N$ scattering, dictating the soft 
$\pi NN(\Delta)$ form factor quoted above~\cite{Rapp:1999ej}. The resulting $\rho$ 
spectral function in nuclear matter is substantially broadened, with a 
(non-Breit-Wigner) shoulder around $M$$\simeq$0.5\,GeV; this is precisely the region 
where most of the free $\omega\to\rho\pi$ decays occur. 
Note that spacelike parts of the $\pi$ and $\rho$ spectral functions (i.e., with negative 
4-momenta squared, $q^2$$<$0) contribute to $\Gamma_{\omega\to\pi\rho}^{\rm med}$; they 
correspond to $t$-channel exchanges in $\omega N$ scattering (e.g., $\rho$ exchange in 
$\omega N\to\pi N^*$). For the pion these are encoded in the Lindhard functions in the 
selfenergy, Eq.~(\ref{piself}). For the $\rho$ they also turn out to be 
dominated by the low-lying $P$-wave 
$\rho$-sobars, $\rho NN^{-1}$ and $\rho\Delta N^{-1}$. The latter is well constrained 
by nuclear photo-absorption ($f_{\rho\Delta N}^2/4\pi$=16.2, 
$\Lambda_{\rho\Delta N}$=0.7\,GeV), but the purely spacelike $NN^{-1}$ mode 
(generating Landau damping of the exchanged $\rho$) is not. An analysis of $\rho$ 
photo-production cross sections, $\gamma p\to \rho p$~\cite{Riek:2008ct}, gave 
indications for a rather soft form factor, $\Lambda_{\rho NN}$$\simeq$0.6\,GeV 
($f_{\rho NN}^2/4\pi$=6.0), but it might be as soft as the $\pi NN$ 
form factor in the pion cloud of the $\rho$. This needs to be investigated in  
future analysis of $\omega N$ scattering data. Here, we bracket the uncertainty 
by varying $\Lambda_{\rho NN}$=0.3-0.6\,GeV and $g'_{NN}$=0-0.6. We find that the 
$\omega$ coupling to spacelike $S$-wave rhosobars (e.g., $N^*(1520)N^{-1}$, 
corresponding to $\omega N\to \pi N^*(1520)$) is already much less 
important.

In addition to modifications of the $\pi\rho$ cloud, pion dressing in the direct 
$\omega\to\pi\pi\pi$ channel and $\omega N^*N^{-1}$ excitations occur. 
The direct 3$\pi$ decay has considerable phase space in vacuum, and thus we
expect its in-medium modification to be smaller than for the $\pi\rho$ channel, 
especially if the latter dominates in vacuum and with our soft form factors 
for the pion dressing; for $\Lambda_{\pi NN(\Delta)}$=0.3\,GeV
we estimate $\Gamma_{\omega\to3\pi}^{\rm med}(\varrho_0)$$<$20\,MeV 
based on recent work in Ref.~\cite{Ramos:2013mda}.
For the $\omega$-sobars, e.g., $N^*(1535)$, $N^*(1520)$ or
$N^*(1650)$~\cite{Lutz:2001mi,Muehlich:2006nn}, we cannot simply adopt the
couplings from the literature, since they were adjusted to fit $\omega N$ scattering 
data without the inclusion of $\pi\rho$ cloud effects. If the latter are present,
the direct-resonance contributions need to be suppressed to still describe
$\omega N$ scattering, and thus their contribution to the in-medium width will
be (much) smaller than in Refs.~\cite{Lutz:2001mi,Muehlich:2006nn}. 

\section{$\omega$ Width in Nuclear Matter}
\label{sec:results}
\begin{figure}[!tb]
\includegraphics[width=0.47\textwidth]{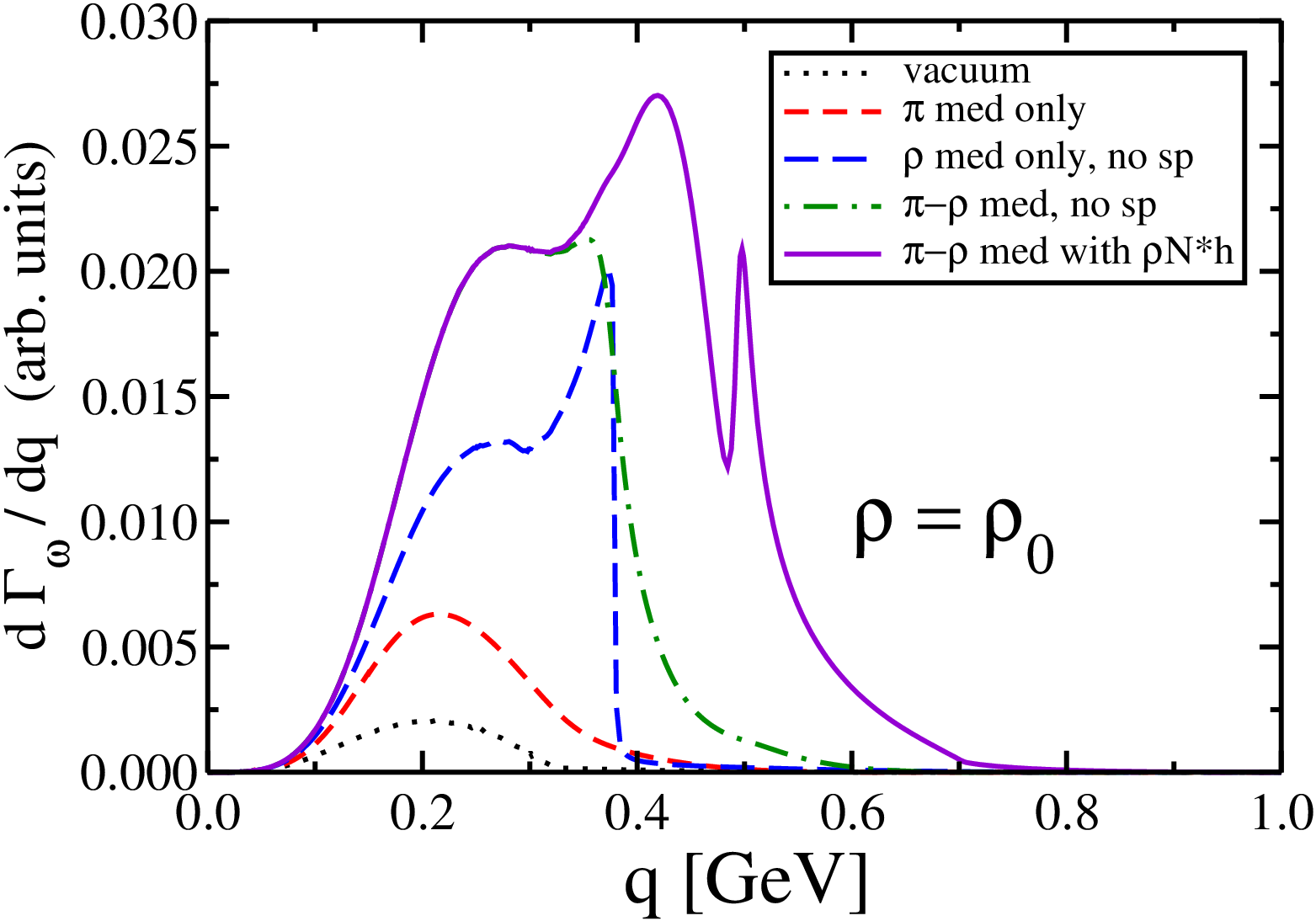}
\hspace{0.5cm}
\includegraphics[width=0.47\textwidth]{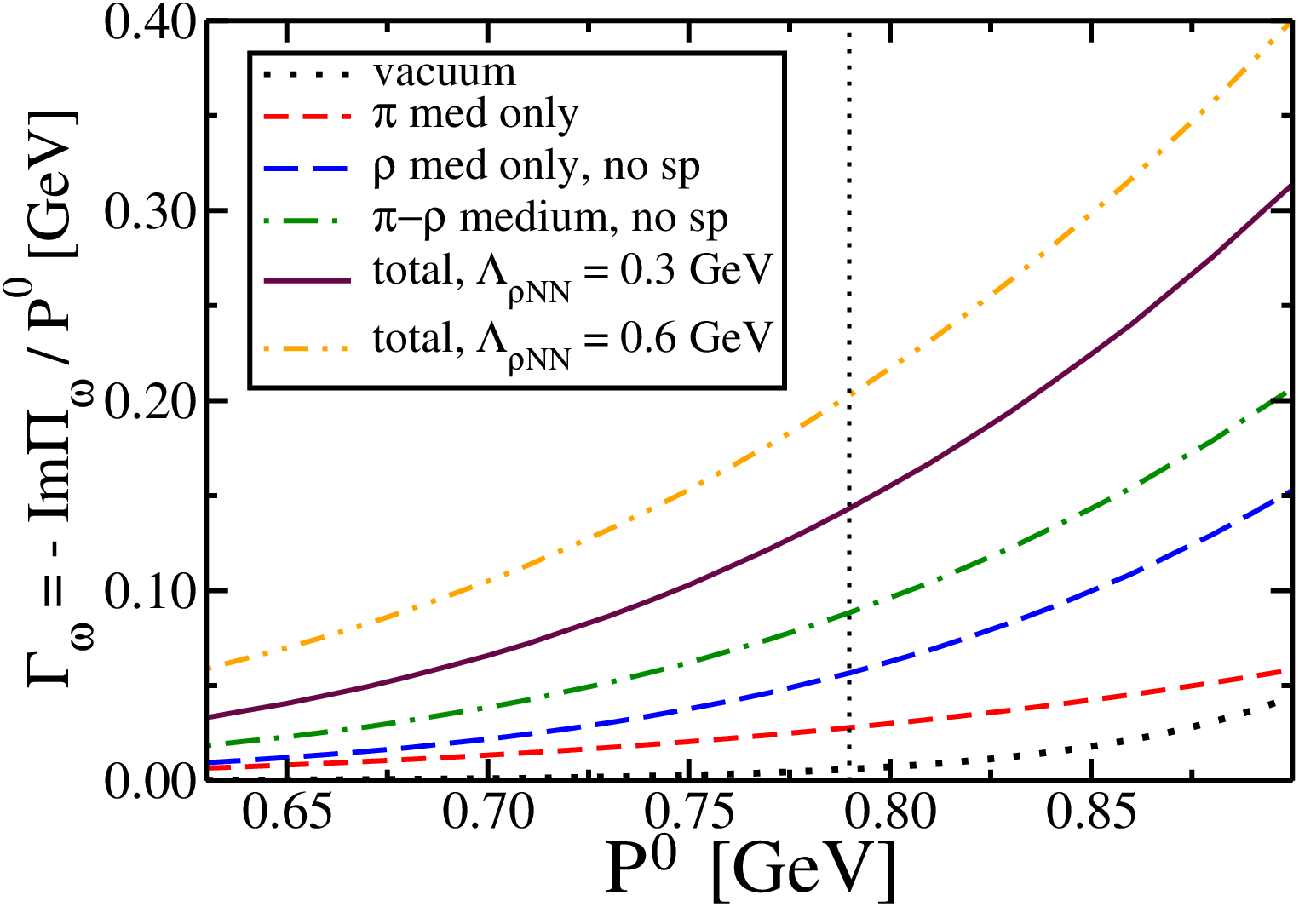}
\caption{Left: differential decay momentum distribution of the $\omega\to\rho\pi$ 
width (for $m_\omega$=782\,MeV) in vacuum (dotted line) and at saturation 
density when dressing either the pion (short-dashed line) or the $\rho$ (long-dashed line), 
or both (dash-dotted line), without spacelike $\rho$ modes. The solid line includes 
spacelike $\rho$'s, where the two maxima beyond $q$$\simeq$0.4\,GeV correspond 
to $\Delta N^{-1}$ and $NN^{-1}$ excitations ($\Lambda_{\rho NN}$=0.3\,GeV). Right: 
Energy dependence of $\Gamma_{\omega\to\rho\pi}$ at saturation density 
for different contributions as in the left panel.}
\label{fig_dGdq}
\end{figure}
Let us first examine the differential distribution of the $\omega$ width, 
$d\Gamma_\omega/dq$, over the center-of-mass decay momentum, $|\vec{q}|$, of the 
$\pi$ and $\rho$ spectral functions, recall Eq.~(\ref{Pi-omg-spec}). In vacuum, 
the fixed pion mass 
uniquely determines the (off-shell) $\rho$ mass ($M$) at given $q$.  The maximum of 
the distribution occurs at $q_{\rm max}$$\simeq$0.2\,GeV, corresponding to 
$M$$\simeq$0.5\,GeV (see Fig.~\ref{fig_dGdq} left). Consequently, the enhancement 
of the in-medium $\rho$ spectral function around this mass strongly increases the 
phase space and thus $\Gamma_{\omega\to\pi\rho}^{\rm med}$. A similar, albeit less 
pronounced effect is caused by the in-medium pion. A further remarkable increase in 
decay width is generated by spacelike $\rho$-sobars above $q$$\simeq$0.4\,GeV, 
which, for a free pion ($m$=$m_\pi$), marks the $M$=0 boundary.
The low-lying collective excitations are sensitive 
to the $\rho NN$ form factor. For a conservative choice of 
$\Lambda_{\rho NN}$=0.3\,GeV, about 40\% of the in-medium $\omega$ width is 
generated by the spacelike $\rho$ modes.

\begin{figure}[!t]
\includegraphics[width=0.47\textwidth]{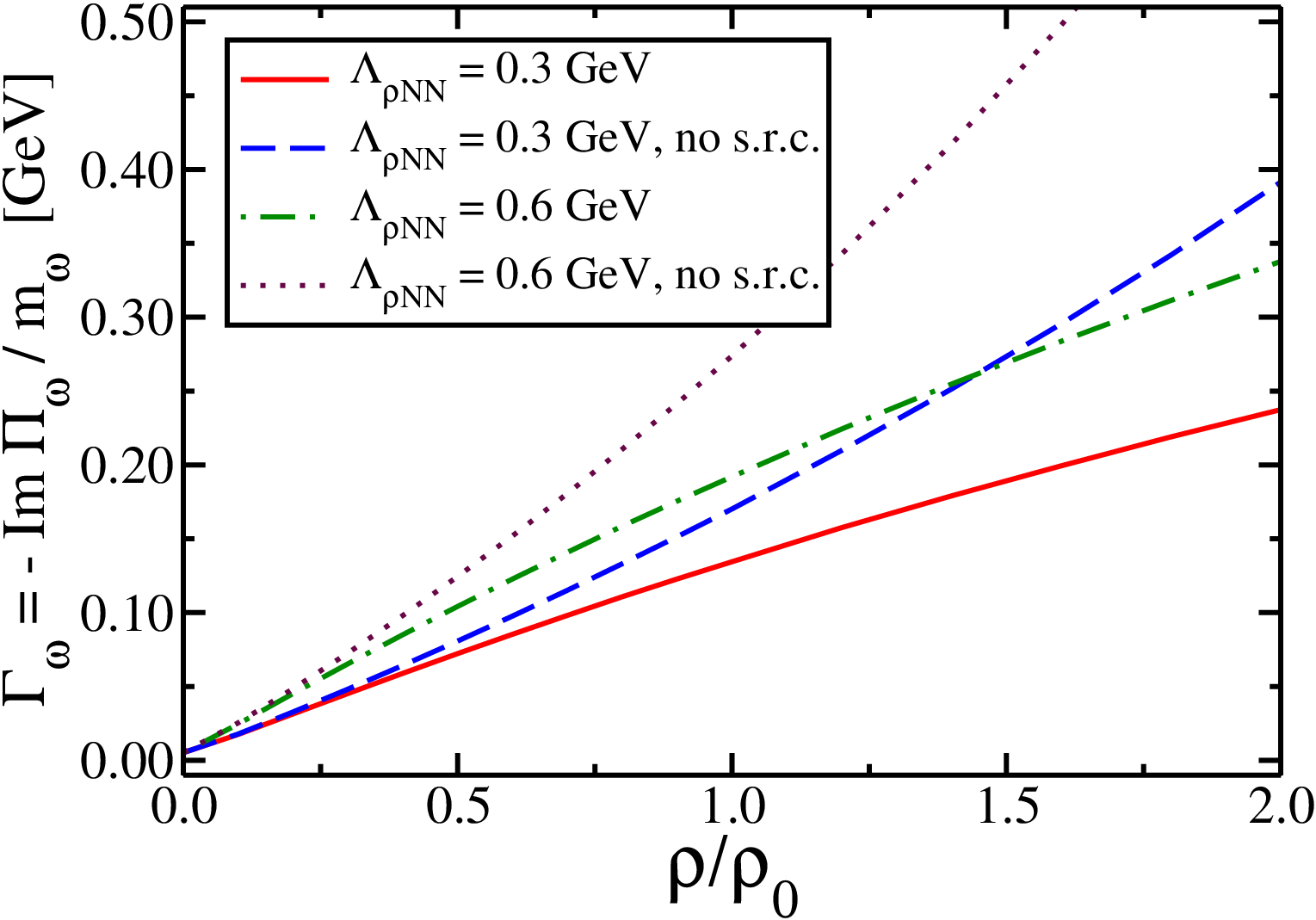}
\hspace{0.2cm}
\includegraphics[width=0.47\textwidth]{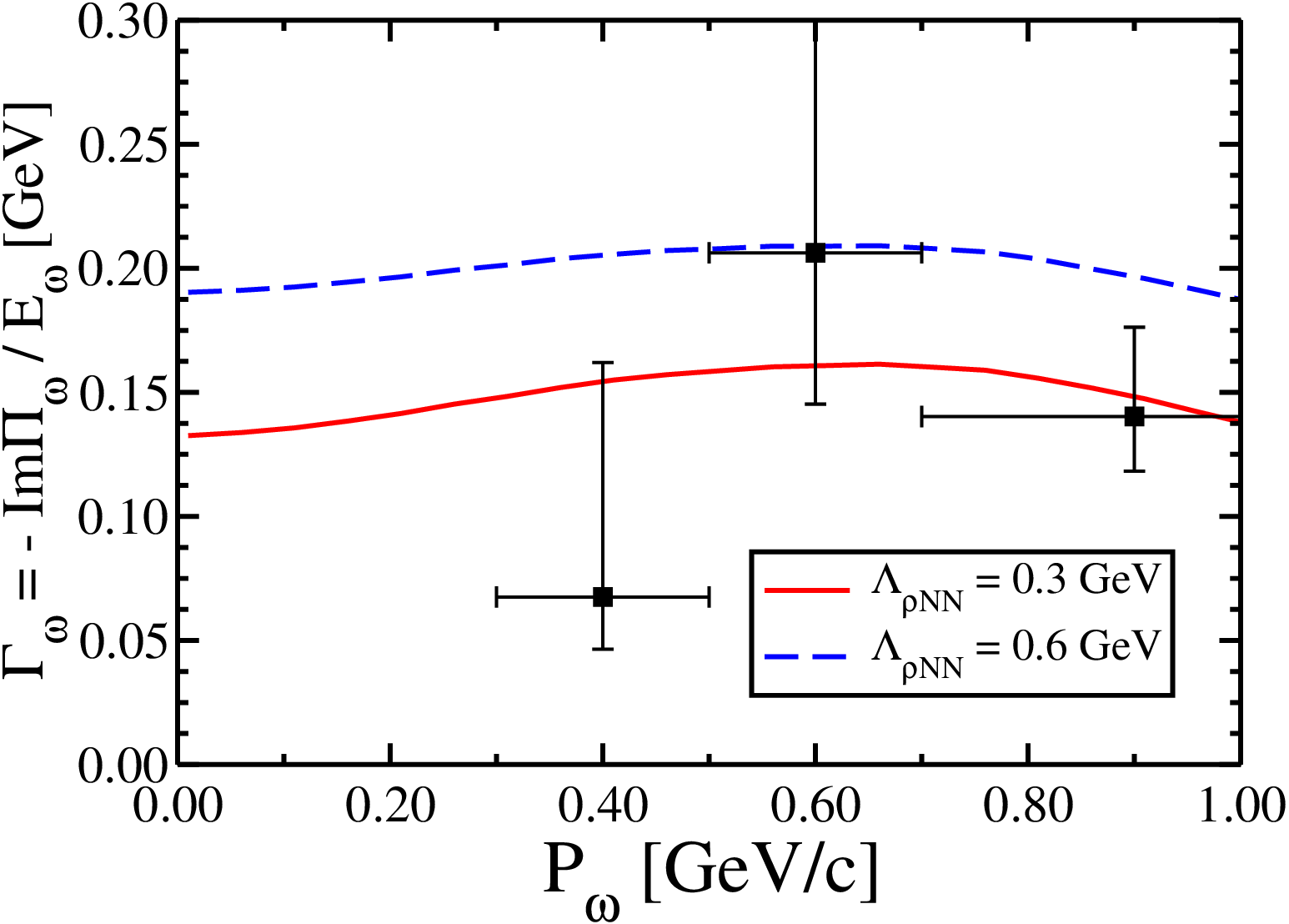}
\caption{Left: Density dependence of the $\omega\to\rho\pi$ width at $P^0=m_\omega$, 
$\vec{P}=\vec{0}$, and for different $\rho NN$ 
form factors and short-range correlations. Right: Three-momentum dependence of 
$\Gamma_{\omega\to\rho\pi}$ at saturation density for on-shell $\omega$ mesons 
($P^2=m_{\omega}^2$, i.e., $E_\omega^2=m_\omega^2+P_\omega^2$), 
compared to CBELSA/TAPS data~\cite{Kotulla:2008aa}.}
\label{fig_gam}
\end{figure}

The energy dependence of $\Gamma_{\omega\to\pi\rho}^{\rm med}$ is rather pronounced
(Fig.~\ref{fig_dGdq} right), a remnant of the (nominal) vacuum $\pi\rho$ threshold together 
with the $\vec{q}\,^2$ dependence of the $\omega\pi\rho$ vertex. The
density dependence of $\Gamma_{\omega\to\pi\rho}^{\rm med}$ (Fig.~\ref{fig_gam}
left) exhibits significant nonlinearities. At normal nuclear matter density, the
dominant uncertainty is due the $\rho NN$ form factor, quantified as
$\Gamma_{\omega\to\pi\rho}^{\rm med}$=130-200\,MeV.

The 3-momentum dependence of the on-shell $\omega$ width (i.e., for 
$P^2$=$(P^0)^2$$-$$\vec{P}\,^2$=$m_{\omega}^2$), relative to the nuclear rest frame, 
turns out to be moderate (Fig.~\ref{fig_gam} right), as generally expected from cloud 
effects with soft formfactors counter-acting the momentum dependence of the 
vertices. 
A fair agreement with CBELSA/TAPS data~\cite{Kotulla:2008aa} is found, apparently 
preferring the lower values of $\Lambda_{\rho NN}$, leaving room for 
(smaller) contributions from direct 3$\pi$ and interference terms, as well as from
$\omega$-sobars which are expected to come in at higher 3-momenta~\cite{Muehlich:2006nn}. 
However, we recall the somewhat larger in-medium width of $\sim$200\,MeV found by 
CLAS~\cite{Wood:2010ei}.

In the very recent work of Ref.~\cite{Ramos:2013mda}, the total $\omega$ width in nuclear 
matter is computed with similar methods.
At $\varrho_N$=$\varrho_0$ and $\vec{P}=\vec{0}$, $\Gamma_\omega^{\rm med}=129\pm10$~MeV 
is reported, predominantly due to the $\rho\pi$ cloud modification and with a more 
pronounced momentum dependence. 
The $\rho$ spectral function employed in there exhibits a factor of $\sim$2 less
broadening than in our input, while the pion modifications are stronger due to a harder
$\pi NN$ formfactor. We recall that the latter is fixed in our approach as part of the 
quantitatively constrained $\rho$ spectral function. It was also argued in 
Ref.~\cite{Ramos:2013mda} that medium effects in interference terms of 3$\pi$ final 
states from direct 3$\pi$ and $\rho\pi$ decays, which we neglected here, are small.
Thus both our work and Ref.~\cite{Ramos:2013mda} identify the $\pi\rho$ cloud as
the main agent for the $\omega$'s in-medium broadening, albeit with some differences
in the partitioning into $\pi$ and $\rho$ modifications, and in the 3-momentum
dependence.  

\section{Summary}
\label{sec:sum}
We have studied the width of the $\omega$ meson in cold nuclear matter focusing
on the role of its $\pi\rho$ cloud. We have employed hadronic many-body theory 
utilizing pion and $\rho$ propagators evaluated with the same techniques,
constrained and applied previously in both elementary and heavy-ion reactions.
The low-mass shoulder in the in-medium $\rho$ spectral function, together with spacelike 
contributions in the $\pi\rho$ intermediate states, induce large effects, along with 
non-linear density dependencies, not captured in previous calculations based on 
$T$-$\varrho$ approximations. For an $\omega$ at rest at saturation density, we find
$\Gamma_\omega^{\rm med}$=130-200\,MeV, where the uncertainty is largely due to the 
$\rho NN$ vertex formfactor which could not be accurately constrained before from 
$\rho$ properties alone. Together with a rather weak 3-momentum dependence of the 
on-shell $\omega$ width, our calculations compare favorably with data from recent 
absorption experiments. The present uncertainties can be reduced by systematic analyses 
of vacuum $\omega$ scattering data (similar to the $\pi NN$ form factor in the $\rho$ 
cloud), where also contributions from direct 3$\pi$ couplings and $\omega N$ resonances 
($\omega$-sobars) need to be included. Work in this direction is in 
progress. The emergence of a large $\omega$ width from $\rho$ and pion propagators 
in nuclear matter is encouraging, and corroborates the quantum many-body approach 
as a suitable tool to assess the properties of hadrons in medium.

\vspace{0.5cm}

\noindent 
{\bf Acknowledgment}\\
This work has been supported by the U.S. National Science Foundation
under grant no.~PHY-1306359,
the Humboldt Foundation, 
the BMBF (Germany) under project no.~05P12RFFCQ,
the Ministerio de Econom\'{\i}a y Competitividad (Spain) under grant 
FPA2011-27853-C02,  the Centro Nacional de F\'isica de Part\'iculas, Astropart\'iculas 
y Nuclear (Consolider-Ingenio 2010) and the EU Integrated Infrastructure Initiative 
Hadron Physics Project under Grant Agreement no.~227431.

\end{document}